\documentclass[11pt]{article}
\usepackage{amsmath,amsfonts,amssymb,amsthm,verbatim,color,lscape}
\usepackage[figuresright]{rotating}
\usepackage[sort&compress]{natbib}
\begin{document}
\def\a{\alpha}
\newcommand{\bea}{\begin{eqnarray}}
\newcommand{\eea}{\end{eqnarray}}

\def\R{\mathbf{R}}
\def\I{\mathbf{I}}
\def\x{\mathbf{x}}
\def\X{\mathbf{X}}
\def\bfv{\mathbf{v}}
\def\V{\mathbf{V}}
\def\Y{\mathbf{Y}}
\def\y{\mathbf{y}}
\def\bbf{\boldsymbol{\beta}}
\def\gbf{\boldsymbol{\gamma}}
\def\obf{\boldsymbol{\omega}}
\def\Obf{\boldsymbol{\Omega}}
\def\U{\mathbf{U}}
\def\J{\mathbf{J}}
\def\ga{\gamma}
\def\u{\mathbf{u}}
\def\C{\mathbf{C}}
\def\p{\mathbf{p}}
\def\1{\mathbf{1}}
\def\t{\mathbf{t}}
\def\W{\mathbf{W}}

\title{On the estimation of normal copula discrete regression models using the continuous extension and simulated likelihood}

\date{}
\author{
Aristidis K. Nikoloulopoulos\footnote{{\small\texttt{A.Nikoloulopoulos@uea.ac.uk}}, School of Computing Sciences, University of East Anglia,
Norwich NR4 7TJ, UK}
}
\maketitle

\vspace{2ex}

\begin{abstract}
\noindent
The continuous extension of a discrete random variable is amongst the computational methods used for estimation of multivariate normal copula-based models with discrete margins. Its advantage is that the likelihood can  be derived conveniently under the theory for copula models with continuous margins, but there has not been a clear analysis of the adequacy of this method. We investigate the asymptotic and small-sample efficiency of two variants of the method  for estimating the multivariate normal  copula with univariate binary, Poisson, and negative binomial regressions, and  show that they lead to biased estimates for the latent correlations, and the univariate marginal parameters that  are not regression coefficients. We implement  a maximum simulated  likelihood  method, which is based  on evaluating the multidimensional integrals of the likelihood with  randomized quasi Monte Carlo methods. Asymptotic and small-sample efficiency calculations show that our method is nearly as efficient as maximum likelihood for fully specified multivariate normal copula-based models. An illustrative example is given to show the use of our simulated likelihood method.
\\\\
\noindent {\it Keywords:} {Continuous extension; Jitters; Multivariate normal copula; Rectangle probabilities; Simulated likelihood.}
\end{abstract}

\maketitle

\section{Introduction}
For multivariate discrete data $\mathbf Y=(Y_1, \ldots , Y_d)$ given a vector of (continuous or discrete) covariates
$\mathbf{x}=(\mathbf{x}_1,\ldots,\mathbf{x}_d)$ with $\x_j\in \mathbb{R}^p,\,j=1,\ldots,d$, the discretized multivariate normal (MVN) distribution, or the MVN copula with discrete margins,
has been in use for a considerable length of time, e.g.  \cite{joe97}, and much earlier in the biostatistics \citep{Ashford&Sowden1970}, psychometrics \citep{Muthen1978}, and econometrics \citep{Hausman&Wise1978} literature.  It is usually known as a multivariate, or multinomial, probit model. The multivariate probit model is a simple example of the MVN copula with univariate probit regressions as the marginals. In the general case, the discretized MVN model has the following cumulative distribution function (cdf):\begin{equation}\label{MVN}
\Pr (Y_1 \leq y_1, \ldots , Y_d \leq y_d;\mathbf{x})=\Phi_d\left(\Phi^{-1}[F_{Y_1}(y_1;\mathbf{x}_1)],\ldots,
  \Phi^{-1}[F_{Y_d}(y_d;\mathbf{x}_d)];\R\right),
\end{equation}
where $\Phi_d$ denotes the standard MVN distribution function with correlation
matrix $\R=(\rho_{jk}: 1\le j<k\le d)$, $\Phi$ is the cdf of the univariate standard normal,
and $F_{Y_1}(y_1;\mathbf{x}_1),\ldots,$ $F_{Y_d}(y_d;\mathbf{x}_d)$ are the univariate discrete
cdfs.

Other multivariate copulas for discrete response data have been around a long time, e.g. in \cite{joe97}, and earlier for some simple copula models.
Simple parametric families of copulas have a closed form cdf; hence the joint likelihood
is straightforward to derive from the probability mass function (pmf) as a finite difference of the cdf, but  they provide limited dependence.   For example, \cite{meester&mackay94} used a Frank copula \citep{frank79} with a closed form cdf to model multivariate binary data. The Frank copula is a member of the Archimedean class of copulas, which are limited to exchangeable structures.

The MVN copula generated by the MVN distribution inherits the useful properties of the latter, thus
allowing a wide range for dependence, and overcomes the drawback of limited dependence inherent in simple parametric
families of copulas \citep{nikoloulopoulos&joe&chaganty10}.
The use of the MVN with logistic regression (or Poisson or  negative binomial regression) is just a special case of the general theory of dependence modelling with copulas. Implementation of the MVN copula for discrete data (discretized MVN) is possible,
but not easy, because the MVN distribution as a latent model for discrete response requires  rectangle probabilities based on high-dimensional integrations or their approximations  \citep{Nikoloulopoulos&karlis07FNM}.  Many approaches have been considered for computing  high-dimensional normal probabilities, see e.g. \cite{schervish84}, \cite{genz92}, and \cite{joe95}. These could be used to evaluate the normal copula-based likelihood for discrete data  with general dependence.

Four recent papers \citep{heinen&renf07,heinen&renf08,madsen09,Madsen&Fang11} attempt to ``approximate" the likelihood, by using the continuous extension (CE) of a discrete random variable developed in \cite{denuit&lambert05}. \cite{heinen&renf07,heinen&renf08} proposed  a surrogate likelihood, assuming the latent uniform variables in the CE of a discrete random variable are observed, while  \cite{madsen09} and  \cite{Madsen&Fang11} used also the CE but their method is actually an example of a simulated likelihood to compute the MVN rectangle probabilities.
The methods based on the CE cannot be recommended until its properties have been studied and compared to existing  methods.
The CE method in \cite{denuit&lambert05} has been used to prove theoretical results for copula-based concordance measures for discrete data.
Although its application to copula dependence modeling for discrete data   is novel, its theoretical and small-sample efficiency has yet to be established in that context. The contribution in this paper is (a) to  examine thoroughly the accuracy and the adequacy of the surrogate and simulated likelihood method based on the CE using asymptotics and simulations;  (b) to improve the efficiency of simulated likelihood by transforming the rectangle integrals as in \cite{genz92}; and (c) to give precise guidelines for  handling the MVN copula-based likelihood for regression models with dependent discrete response data.

For ease of exposition,  we consider the case that the univariate marginal parameters are common to different univariate margins;  $\gbf$ denotes the $r$-dimensional vector of univariate marginal parameters that are not regression coefficients, and $\bbf$ the $p$-dimensional vector of the regression parameters. The marginal means, $E(Y_j)=\mu_j,\,j=1,\ldots,d$, depend on the vector of the covariates $\x_{j}$, via the  vector $\bbf$  and a link function,  $\eta(\mu_{j})=\x_{j}^\top\bbf,\,j=1,\ldots,d.$ The remainder of the paper proceeds as follows.
Section 2 provides an overview of the surrogate and simulated likelihood method using the CE. In Section  3 we describe appropriate methods for handling the MVN copula-based likelihood for regression models with dependent discrete response data, and propose a maximum simulated  likelihood  method based  on evaluating the multidimensional integrals of the likelihood with  randomized quasi Monte Carlo methods.
Section 4 and Section 5 contain  theoretical  (e.g. asymptotic properties of the estimators) and small sample  efficiency calculations, respectively, to assess  the  accuracy  of the methods. Section 6 presents an application
of the proposed simulated likelihood  to the toenail infection data in \cite{Madsen&Fang11}. We conclude this article with some
discussion, followed by a technical Appendix.

\section{\label{CE} Surrogate and simulated likelihood using the CE}

\cite{denuit&lambert05} proposed a CE of integer-valued random variables to study concordance measures for dependent discrete data.  They associate an integer-valued random variable $Y$ with a jittered random variable $Y^\star$,
 such that, $$Y^\star=Y+(V-1),$$
where $V$ is a uniform random variable in the unit interval and independent of $Y$. $Y^\star$ is a CE of $Y$, and
has support on the reals.
Given a particular realization of the jittering, the (conditional) density of $Y^\star$
is $f_{Y^\star|V}(y^\star|v)=f_Y(\lceil y^\star\rceil),$
where $\lceil y^\star\rceil$ is the ceiling of $y^\star$, or the smallest integer greater than or equal to $y^\star.$
The (conditional) cdf of $Y^\star$ is $F_{Y^\star|V}(y^\star|v)=F_Y(\lceil y^\star\rceil-1) + vf_Y(\lceil y^\star\rceil).$
It is easy to see that the simplified  cdf and density  of $Y^\star$  take the form
$F_{Y^\star|V}(y^\star|v)=F_Y(y-1) + vf_Y(y)$ and $f_{Y^\star|V}(y^\star|v)=f_Y(y),$  respectively.

\subsection{\label{HR-subsec}Surrogate likelihood based on the CE of a discrete random variable}
\cite{heinen&renf07,heinen&renf08} were the first that
adopt this approach to form a surrogate likelihood for MVN copula-based models with discrete margins (hereafter HR method). Copula-based models were originally developed for continuous responses where the density is obtained using partial derivatives of the multivariate cdf (see e.g. \cite{Nikoloulopoulos&joe&li11}),  and hence the numerical calculations are much simpler.
The corresponding density for the jittered continuous vector $\mathbf Y^\star=(Y_1^\star, \ldots , Y_d^\star)$ given $n$ independent standard uniforms $\mathbf V=(V_1,\ldots,V_d)$  is,
$$
h_{\mathbf Y^\star|\mathbf V}(\mathbf{y}^\star|\mathbf{v};\mathbf{x})=\\
c\Bigl(F_{Y_1^\star|V_1}(y_1^\star|v_1;\x_1),\ldots,F_{Y_d^\star|V_d}(y_d^\star|v_d;\x_d);\R\Bigr)\prod_{j=1}^d f_{Y_j^\star|V_j}(y_j^\star|v_j;\x_j),
$$
where $\bfv=(v_1,\ldots,v_d)$ are realizations of the jitters $\V$ and  $c(u_1,\ldots,u_d;\R)$ is the $d$-variate normal copula density.
Since the MVN copula has a closed form density,
$$c(u_1,\ldots,u_d;\R)=|\R|^{-1/2}\exp\Bigl[\frac{1}{2}\bigl\{\mathbf q^\top(\I_d-\R^{-1})\,\mathbf q\bigr\}\Bigr],$$
where $\mathbf q=(q_1,\ldots,q_d)$ with $q_j=\Phi^{-1}(u_j), j=1,\ldots,d$ and $\I_d$ is the $d$-dimensional identity matrix,
the authors avoid the  multidimensional  integration by using the CE of the discrete random variables.
The estimated parameters can be obtained by  maximizing  the surrogate log-likelihood,
$$\ell_{HR}(\bbf,\gbf,\R)\\
= \sum_{i=1}^{n}
\log{h_{\mathbf Y^\star|\mathbf V}(y_{i1}^\star,\ldots,y_{id}^\star|v_{i1},\ldots,v_{id};\x_{i1},\ldots,\x_{id})},$$
over the univariate and copula parameters $(\bbf,\gbf,\R)$.
In order to avoid the noise introduced by the jitters $\mathbf V,$ they use $m$ jitters.
That is,  they simulate a vector of independent standard uniforms  $\mathbf V_k=(V_{1,k},\ldots,V_{d,k})$  and maximize,
$$\ell_{HR}^{(k)}(\bbf,\gbf,\R)
= \sum_{i=1}^{n}
\log{h_{\mathbf Y^\star|\mathbf V_k}(y_{i1}^\star,\ldots,y_{id}^\star|v_{i1,k},\ldots,v_{id,k};\x_{i1},\ldots,\x_{id})},$$ for $k=1,\ldots,m.$
The HR estimates are the average of the estimates over the $m$  runs.

\subsection{\label{MF-subsec}Simulated likelihood based on the CE of a discrete random variable}
\cite{madsen09} and \cite{Madsen&Fang11} also adopt the CE of $Y_j$ proposed by \cite{denuit&lambert05}, but they make use of importance functions in combination with simulated likelihood.
Simulated likelihood for multivariate probit models has been used in econometrics  since the early 1990s, and it is one of the recommended methods for generalized linear mixed models, see e.g. Chapter 7 in \cite{Demidenko04}.
The simulated likelihood method, applies importance sampling to simulate the likelihood function, but this must be done in a way that the simulated likelihood changes little when the parameters of the model are perturbed a little. This can be accomplished by using the same set of random draws in an appropriate way.  For  importance sampling to work well,  the integrand, say,
$$\int {e(z)} \,dz=\int \frac{e(z)}{g(z)}  g(z) \,dz=E_Z\Bigl[\frac{e(z)}{g(z)} \Bigr],$$ must be converted to a good form of an empirical average based on data simulated from a suitable importance sampling distribution $g$.
Better choices of $g$ have small variance for $w(z)=e(z)/g(z)$, such that the importance weight $w$ is bounded.
A bad choice can lead to an inflated variance of the integral-estimate and thus to a very poor approximation.
Hence, in order to use simulated likelihood in an efficient way the choice of the the importance sampling distribution is crucial.

The simulated likelihood using the jitters (hereafter MF method) is,
\begin{equation}\label{MFexp-equation}
L_{MF}(\bbf,\gbf,\R)\\
=E_\mathbf{V}\Bigl[\prod_{i=1}^n h_{\mathbf Y^\star|\mathbf V}(y_{i1}^\star,\ldots,y_{id}^\star|v_{i1},\ldots,v_{id};\x_{i1},\ldots,\x_{id})\Bigr].
\end{equation}
Essentially, the MF importance weight is unbounded. \cite{Madsen&Fang11} write an exponential with quadratic form $\frac{1}{2}\bigl\{\mathbf q^\top(\I_d-\R^{-1})\,\mathbf q\bigr\}$,
which is not positive definite, so for part of the $\mathbf q$ vector space,  it can be
arbitrarily negative; hence their integrand can be arbitrarily large. This is  a
poor choice of an importance function, and it is hard to achieve
 good accuracy for the numerical integral.

The expected likelihood in (\ref{MFexp-equation}) can be approximated by  averaging over the jitters $\mathbf V_k,\,k=1,\ldots,m$, that is,
\begin{multline}\label{MF-equation}
L_{MF}(\bbf,\gbf,\R)=
\frac{1}{m}\sum_{k=1}^m\Bigl[\prod_{i=1}^n h_{\mathbf Y^\star|\mathbf V_k}(y_{i1}^\star,\ldots,y_{id}^\star|v_{i1,k},\ldots,v_{id,k};\x_{i1},\ldots,\x_{id})\Bigr]\\
=\frac{1}{m}\sum_{k=1}^m\left(\prod_{i=1}^n\left\{ |\R|^{-1/2}\exp\left[\frac{1}{2}\left\{\mathbf q_{i,k}^\top(\I_d-\R^{-1})\,\mathbf q_{i,k}\right\}\right]\prod_{j=1}^d f_{Y_j^\star|V_j}(y_{ij}^\star|v_{ij,k};\x_{ij})\right\}\right)\\
=\frac{1}{m}\prod_{i=1}^n\prod_{j=1}^d f_{Y_j}(y_{ij};\x_{ij}) \sum_{k=1}^m\left(\prod_{i=1}^n\left\{ |\R|^{-1/2}\exp\left[\frac{1}{2}\left\{\mathbf q_{i,k}^\top(\I_d-\R^{-1})\,\mathbf q_{i,k}\right\}\right]\right\}\right)\,\,,
\end{multline}
where $\mathbf q_{i,k}=\bigl(\Phi^{-1}[F_{Y_1^\star|V_{1,k}}(y_{i1}^\star|v_{i1,k};\x_{i1})],\ldots,
\Phi^{-1}[F_{Y_d^\star|V_{d,k}}(y_{id}^\star|v_{id,k};\x_{id})] \bigr).$

The MF estimates of $(\bbf,\gbf,\R)$ are  derived by maximizing the $\ell_{MF}=\log L_{MF}$ with respect to  the univariate and copula parameters.
For the
MF (and HR in subsection \ref{HR-subsec})
likelihood, to work well in a numerical optimization routine, the evaluations via simulation have to be smooth (differentiable) when the parameters  change by small amounts. In order to accomplish this, the same set of uniform random variables should be used no matter the parameter values in the iterative optimization; see e.g. \cite{Bhat-Sidharthan-2011}.

\section{\label{ml-ssec} Appropriate methods for handling the likelihood}
Estimation of the model parameters $(\bbf,\gbf,\R)$ can be approached by the standard
maximum likelihood (ML) method, by maximizing the joint log-likelihood \citep{joe97},
\begin{equation}\label{MLlik}
\ell(\bbf,\gbf,\R)\\
= \sum_{i=1}^{n}
\log{h_\mathbf{Y}(y_{i1},\ldots,y_{id};\x_{i1},\ldots,\x_{id}}),
\end{equation}
over the univariate  and copula parameters  $(\bbf,\gbf,\R)$, where $h_\mathbf{Y}$ is the
joint pmf of the multivariate discrete response vector  $\mathbf Y=(Y_1, \ldots , Y_d)$.
\cite{song07}, influencing other authors (e.g. \cite{heinen&renf07,heinen&renf08,madsen09,Madsen&Fang11}), acknowledged  that the pmf can be obtained  as a finite difference of the cdf in (\ref{MVN}).
Generally speaking, this is an imprecise statement, since calculating the finite difference among $2^d$ numerically computed orthant probabilities  may result
in negative values. The pmf can be alternatively  obtained by computing  the following rectangle probability,
\begin{eqnarray}\label{MVNpmf}
h_\mathbf{Y}(\y;\mathbf{x})&=&\Pr (Y_1 = y_1, \ldots , Y_d = y_d;\mathbf{x})\\
&=&\Pr(y_1-1< Y_1\leq y_1,\ldots,y_d-1< Y_d\leq y_d;\mathbf{x})\nonumber\\
&=&\int_{\Phi^{-1}[F_{Y_1}(y_1-1;\mathbf{x}_1)]}^{\Phi^{-1}[F_{Y_1}(y_1;\mathbf{x}_1)]}\cdots
\int_{\Phi^{-1}[F_{Y_d}(y_d-1;\mathbf{x}_d)]}^{\Phi^{-1}[F_{Y_d}(y_d;\mathbf{x}_d)]}  \phi_\R(z_1,\ldots,z_d) dz_1\ldots dz_d,\nonumber
\end{eqnarray}
where $\phi_\R$ denotes the standard MVN density  with latent correlation
matrix $\R$.

The computation of MVN rectangle probabilities such as the one in (\ref{MVNpmf}) is possible,  but requires multidimensional integration. However, there is a special case overlooked in the biostatistics  literature \citep{Kiefer1982,Ochi&Prentice1984,song09,Madsen&Fang11}:
for positive exchangeable correlation structures, the $d$-dimensional integrals
conveniently reduce to 1-dimensional integrals \citep[p. 48]{Johnson&Kotz72}. Hence,
MVN  rectangle probabilities can be quickly
computed to a desired accuracy that is $10^{-6}$ or less, because
1-dimensional numerical integrals are computationally easier than
higher-dimensional numerical integrals.
For general correlation structures,    there are several papers in the literature that focus on the
computation of the MVN rectangle probabilities,  and, conveniently,  the implementation of the proposed algorithms is available in contributed R packages
\footnote{Both approximations to MVN rectangle in \cite{joe95}, the 1-dimensional integral in the exchangeable case, and the method in \cite{schervish84}, can be computed with the functions {\it mvnapp}, {\it exchmvn}, and {\it pmnorm}, respectively, in the R package {\it mprobit} \citep{Joe-Wei-11}. The methods in \cite{genz92}  can be computed with the function {\it pmvnorm} in the R package {\it mvtnorm} \citep{genz-etal-2012}.}.
\cite{schervish84} proposed a locally adaptive numerical integration
method but this method, while more accurate, is  time consuming and restricted to a low dimension. Therefore, \cite{genz92} proposed a  Monte Carlo method and  \cite{joe95} proposed
two approximations to multivariate normal probabilities.  The first-order approximation makes
use of all of the univariate and bivariate marginal probabilities,
and the second-order approximation also makes use of trivariate
and four-variate marginal probabilities.
These advances in computation of MVN probabilities can be used to implement MVN copula models with discrete response data:
\begin{itemize}
\itemsep=0pt

\item
For positive exchangeable dependence structures, if one computes the rectangle MVN probabilities  in (\ref{MLlik}) with the 1-dimensional integral method in \cite{Johnson&Kotz72}, then one is using a numerically accurate likelihood method that is valid for any dimension.
\item If one computes the rectangle MVN probabilities  in (\ref{MLlik}) with the methods in \cite{joe95}, then one is using an approximate likelihood method; see  e.g. \cite{joe97}.
\item If one computes the rectangle MVN probabilities  in (\ref{MLlik}) via simulation based on the  method in \cite{genz92}, then one is using the simulated likelihood method but with something much better than in \cite{madsen09} and \cite{Madsen&Fang11}.
\end{itemize}
Since both approximations in \cite{joe95} are better when the correlations are smaller, we concentrate on the simulated likelihood method based
on the  computation of the  MVN probabilities ala \cite{genz92}.

\subsection{Simulated likelihood method via the optimized method of \cite{genz&bretz02}}
For an integral in the context of an
MVN rectangle probability,
\begin{equation}\label{rect-formula}
P(a_j<Z_j<b_j, j=1,...d)=\int_{a_1}^{b_1}\cdots\int_{a_d}^{b_d}  \phi_\R(z_1,\ldots,z_d) dz_1\ldots dz_d,
\end{equation}
for an MVN density $\phi_\R$ with correlation matrix $\R,$ \cite{genz92} uses a sequence of thee transformations to transform the original integral into an integral over a unit hypercube,
$$
P(a_j<Z_j<b_j, j=1,...d)=\int_{0}^{1}\cdots\int_{0}^{1} e(v_1,\ldots,v_{d-1}) dv_1\ldots dv_{d-1},
$$
with,
\begin{eqnarray*}
e(v_1,\ldots,v_{d-1})&=&e_1e_2(v_1)e_3(v_1,v_2)\ldots e_d(v_1,\ldots,v_{d-1}),\\ e_j(v_1,\ldots,v_{j-1})&=&\varepsilon_j(v_1,\ldots,v_{j-1})-\epsilon_j(v_1,\ldots,v_{j-1}),\\
\epsilon_j(v_1,\ldots,v_{j-1})&=&\Phi\Bigl(\Bigl[a_j-\sum_{k=1}^{j-1}c_{jk}\Phi^{-1}\{\epsilon_k(v_1,\ldots,v_{k-1})+ u_k e_k(v_1,\ldots,v_{k-1})\}\Bigl]/c_{jj}\Bigl),\\
\varepsilon_j(v_1,\ldots,v_{j-1})&=&\Phi\Bigl(\Bigl[b_j-\sum_{k=1}^{j-1}c_{jk}\Phi^{-1}\{\epsilon_k(v_1,\ldots,v_{k-1})+ u_k e_k(v_1,\ldots,v_{k-1})\}\Bigl]/c_{jj}\Bigl);
\end{eqnarray*}
 $\C=(c_{jk}: 1\le j<k\le d)$ is the matrix used for the Cholesky decomposition of $\R$.

This sequence of transformations  reduces the number of integration variables by one, but, more interestingly, the rectangle integral is converted to a bounded integrand, so that  the rectangle probability can be successfully evaluated via importance sampling based on a $(d-1)$-variate   standard uniform random sample $\V_k$,
$$
P(a_j<Z_j<b_j, j=1,...d)=m^{-1}\sum_{k=1}^m e(\bfv_k).
$$

\cite{genz&bretz02} improve the performance of the crude Monte Carlo methods in \cite{genz92} by calling a randomized quasi Monte Carlo method with the use of antithetic variates. They use approximations of the form,
\begin{equation}\label{QMC}
P(a_j<Z_j<b_j, j=1,...d)=m^{-1}\sum_{k=1}^m \frac{1}{2P}\sum_{p=1}^P
(e(|2\lfloor\p_p+\bfv_k\rfloor-\1|) +e(\1-|2\lfloor\p_p+\bfv_k\rfloor-\1|).
\end{equation}
In this form, $\lfloor\t\rfloor$ denotes the vector obtained by taking the fractional part of each of the components of $\t$,  and $\p_p,p=1,\ldots,P$ is a set of quasi-random points.

To sum up, \cite{genz&bretz02} achieve error reduction of Monte Carlo methods with variance reduction methods as (a) transforming to a bounded integrand, (b) using antithetic variates, and (c) using a randomized quasi Monte Carlo method. The test results in \cite{genz&bretz02,genz&bretz2009} show that
their method is very efficient, compared to other methods in the literature.
Note in passing that the method in \cite{genz&bretz02} is ``optimized" in the {\tt mtvnorm} R package \citep{genz-etal-2012}. Hence, on  the calculation of the approximation in (\ref{QMC}),  one doesn't  need to worry about the selection, for example, of the number of jitters $m$, or the number of quasi points $P$.

We implement  a simulated likelihood (hereafter SL), where the rectangle MVN probabilities are computed based on the method in \cite{genz&bretz02}. Since the estimation of the parameters of the MVN copula-based models
is obtained using a quasi-Newton routine \citep{nash90} applied to the log-likelihood in (\ref{MLlik}),  the use
of randomized quasi Monte Carlo simulation to  four decimal place
accuracy for evaluations of integrals works poorly, because
numerical derivatives of the log-likelihood with respect to
the parameters are not smooth. In order to achieve smoothness, the same set of uniform random variables should be used for every rectangle probability that comes up in the optimization of the SL.
Hence, our implementation allows estimation of parameters from response vectors of dimension much larger than three  that used in previous theoretical studies and applications (e.g. \cite{song07,song09}).

\section{Theoretical efficiency}
In this section, we perform several theoretical calculations, similarly to  \cite{joe95,joe08},  to
investigate the accuracy of the  ``approximate" likelihood methods   \citep{heinen&renf07,heinen&renf08,madsen09,Madsen&Fang11}, and the SL method in Subsection 3.1, which is based  on evaluating the multidimensional integrals at the likelihood with the method in \cite{genz&bretz02}.

\subsection{Asymptotics}
In this subsection, we study the asymptotics of the HR and  SL methods, and we assess the accuracy based on the limit (as the number of clusters
increases to infinity) of the maximum surrogate likelihood estimate  (HRMLE) and the maximum SL estimate  (MSLE).
By varying factors such as dimension $d$, regression and not regression parameters, the amount of discreteness (binary versus count response), and latent correlation  for exchangeable structures,   we demonstrate patterns
in the asymptotic bias of the HRMLE and MSLE, and assess the performance of HR and SL.
For the cases where we compute the probability limit, we will take a constant dimension $d$ that increases. We  will also conveniently use discrete covariates so that we can assume that there are a finite number of distinct values. The idea of the continuous extension is to replace a numerically more difficult MVN rectangle probability calculation with a simpler MVN density value, and hence it is discrete responses that matter and not the type of covariates. The pattern should be similar with continuous covariates but the bias cannot be determined as easily.

Let the $T$ distinct cases for the discrete response and the covariates be denoted as
$$(\y^{(1)},\x^{(1)}),\ldots, (\y^{(T)},\x^{(T)}),$$
where $\y^{(t)}=(y_1^{(t)},\ldots,y_d^{(t)}),\,\x^{(t)}=(\x_1^{(t)},\ldots,\x_d^{(t)}),\, t=1,\ldots, T.$
In a random sample of size $n$, let the corresponding frequencies be denoted as $n^{(1)},\ldots, n^{(T)}$. Assuming a probability
distribution on the covariates, for $t = 1,\ldots, T$, let $p^{(t)}$ be the limit in probability of $n^{(t)}/n$ as $n\to \infty$.
For the simulated likelihood in (\ref{MLlik}), we have the limit,
\begin{equation}\label{limitslik}
n^{-1}\ell(\bbf,\gbf,\rho)\to \sum_{t=1}^{T} p^{(t)}
\log{h_\mathbf{Y}(y_{1}^{(t)},\ldots,y_{d}^{(t)};\x_{1}^{(t)},\ldots,\x_{d}^{(t)}}),
\end{equation}
where $h_\mathbf{Y}(\mathbf y^{(t)};\mathbf{x}^{(t)})$ is computed using the method in \cite{genz&bretz02}.
The limit of the  MSLE (as $n\to \infty$) is the maximum of (\ref{limitslik});
we denote this limit as $(\bbf^{SL},\gbf^{SL},\rho^{SL})$. Note in passing that the limit of the  standard MLE (as $n\to \infty$) is the maximum of (\ref{limitslik}) where $h_\mathbf{Y}(\mathbf y^{(t)};\mathbf{x}^{(t)})$ is computed  with the 1-dimensional integral method in \cite{Johnson&Kotz72}.

The surrogate log-likelihood based on the MVN
density with exchangeable dependence structure is,
$$\ell_{HR}(\bbf,\gbf,\rho)=\sum_{i=1}^n\log\Bigl[c(u_{i1},\ldots,u_{id};\rho)\prod_{j=1}^d f_{Y_j}(y_{ij};\x_{ij})\Bigr],$$
where
$c(u_{i1},\ldots,u_{id};\rho)=\frac{1}{\sqrt{[1+(d-1)\rho](1-\rho)^{d-1}}}
e^{\frac{-\rho}{2(1-\rho)[1+(d-1)\rho]}\bigl((d-1)\rho\sum_{j=1}^d q_{ij}^2-2\sum_{j<k}q_{ij}q_{ik}\bigr)}$
\citep{zezula09} with $q_{ij}=\Phi^{-1}(u_{ij})=\Phi^{-1}[F_{Y_j}(y_{ij}-1;\x_{ij}) + v_{ij} f_{Y_j}(y_{ij};\x_{ij})]$ as realizations of standard normal random variables.
Therefore $n^{-1}\ell_{HR}(\bbf,\gbf,\rho)$ is,
\begin{eqnarray*}
&&-\frac{1}{2}\log[1+(d-1)\rho]-\frac{d-1}{2}\log(1-\rho)-n^{-1}\frac{\rho}{2(1-\rho)[1+(d-1)\rho]}\times\\
&&\Bigl[(d-1)\rho\sum_{i=1}^n\sum_{j=1}^d q_{ij}^2-2\sum_{i=1}^n\sum_{j<k}q_{ij}q_{ik}\Bigr]+n^{-1}
\sum_{i=1}^n\sum_{j=1}^d \log f_{Y_j}(y_{ij};\x_{ij}).
\end{eqnarray*}
Then as $n\to\infty,$ the limit in probability of $n^{-1}\ell_{HR}(\bbf,\gbf,\rho)$ is,
\begin{eqnarray}\label{limitlik}
&&\sum_{t=1}^{T} p^{(t)} \times \Bigl[-\frac{1}{2}\log[1+(d-1)\rho]-\frac{d-1}{2}\log(1-\rho)-\frac{\rho}{2(1-\rho)[1+(d-1)\rho]}\times\nonumber\\
&&\bigr\{(d-1)\rho\sum_{j=1}^d \xi_{j}^{(t)}-2\sum_{j<k}\zeta_{j}^{(t)}\zeta_{k}^{(t)}\bigr\}+
\sum_{j=1}^d \log f_{Y_j}(y_{j}^{(t)};\x_j^{(t)})\Bigr],
\end{eqnarray}
where $\zeta_{j}^{(t)},\xi_{j}^{(t)},\,j=1,\ldots,d, t=1,\ldots,T$ are conditional expectations for the truncated normal distribution that have closed forms. Further details are given in the Appendix.
The limit of the  HRMLE (as $n\to \infty$) is the maximum of (\ref{limitlik});
we denote this limit as $(\bbf^{HR},\gbf^{HR},\rho^{HR})$.

We will compute these limiting  HRMLE and MSLE in a variety of situations to show clearly if  the HR and SL methods
are good. By using these limits, we do not need Monte Carlo simulations for comparisons, and we can quickly vary parameter
values and see the effects.
The $p^{(t)}$
in (\ref{limitslik}) and (\ref{limitlik}) are the model based probabilities $h_\mathbf{Y}(\mathbf y^{(t)};\mathbf{x}^{(t)})$, and computed with the 1-dimensional integral method in \cite{Johnson&Kotz72}.
For marginal models we use Bernoulli$(\mu)$, Poisson$(\mu)$, and  negative binomial (NB). For the latter model, we use both the NB$1(\mu,\,\ga)$  and NB$2(\mu,\,\ga)$ parametrization in \cite{cameron&trivedi98}; the NB$2$ parametrization is that used in \cite{lawless87}. For a count response, we get a finite number of $\y^{(t)}$ vectors by truncation. The truncation point  is chosen to exceed $0.999$ for total probabilities. Further, we use only one binary covariate, which is the same for each cluster; this scenario is typical for longitudinal
data with time-independent covariates. Other discrete (and time-dependent) covariates can be used,  but computations are more time consuming because $T$ is larger.

\begin{table}[!h]
\begin{small}
\begin{center}
\begin{tabular}{ccccccc}
\hline
         $d$ &        $\rho$ &   $\rho^{HR}$ &        $\beta_0$ &   $\beta_0^{HR}$          &        $\beta_1$ &   $\beta_1^{HR}$    \\
\hline
         2 &        0.3 &      0.120 &       -0.5 &     -0.499 &        0.5 &      0.499 \\

         2 &        0.6 &      0.255 &       -0.5 &     -0.497 &        0.5 &      0.497 \\

         2 &        0.8 &      0.368 &       -0.5 &     -0.493 &        0.5 &      0.493 \\
\hline
         5 &        0.3 &      0.120 &       -0.5 &     -0.498 &        0.5 &      0.498 \\

         5 &        0.6 &      0.255 &       -0.5 &     -0.491 &        0.5 &      0.491 \\

         5 &        0.8 &      0.368 &       -0.5 &     -0.479 &        0.5 &      0.479 \\
\hline
        10 &        0.3 &      0.120 &       -0.5 &     -0.496 &        0.5 &      0.496 \\

        10 &        0.6 &      0.255 &       -0.5 &     -0.484 &        0.5 &      0.484 \\

        10 &        0.8 &      0.369 &       -0.5 &     -0.467 &        0.5 &      0.467 \\
\hline
\end{tabular}
\caption{\label{HR-logistic-bias} Limiting HRMLE for  MVN copula-based models with marginal logistic regression.}
\end{center}
\end{small}
\end{table}

Representative results are shown in Tables \ref{HR-logistic-bias} and \ref{HR-nb-bias} for logistic and NB2 regression, with MSLE results omitted because they were identical with MLE
up to three or four decimal places. Therefore, the SL method leads to unbiased estimating equations.
Regarding the HR method, by varying the latent correlation $\rho$ and dimension $d,$ results are similar for the  binary and count responses. There is substantial asymptotic downward bias for the HRMLE of the latent correlation ($\rho^{HR}$), and it seems that there is negligible asymptotic bias  for the HRMLE for the parameters that  are regression coefficients ($\beta_0^{HR},\beta_1^{HR}$); note that this slightly increases as either $d$ or $\rho$ increases.
Calculating the limit for HRMLE of $\gamma$ ($\gamma^{HR}$) for NB2 regression, it can also be seen that there is substantial asymptotic downward bias for the univariate marginal parameters that  are not regression coefficients as the latent correlation $\rho$ increases. The results in Tables \ref{HR-logistic-bias} and \ref{HR-nb-bias} show that the HR method is adequate with regard to the univariate marginal parameters that  are regression coefficients.

\begin{table}[!h]
\begin{small}
\begin{center}
\begin{tabular}{ccccccccc}
\hline
         $d$ &        $\rho$ &    $\rho^{HR}$        &        $\beta_0$ &   $\beta_0^{HR}$          &        $\beta_1$ &   $\beta_1^{HR}$
         &  $\gamma$ &   $\gamma^{HR}$        \\\hline

         2 &        0.3 &      0.191 &       -0.5 &     -0.498 &        0.5 &      0.495 &        0.5 &      0.480 \\

         2 &        0.6 &      0.397 &       -0.5 &     -0.492 &        0.5 &      0.483 &        0.5 &      0.410 \\

         2 &        0.8 &      0.550 &       -0.5 &     -0.481 &        0.5 &      0.466 &        0.5 &      0.302 \\
\hline
         3 &        0.3 &      0.191 &       -0.5 &     -0.497 &        0.5 &      0.492 &        0.5 &      0.468 \\

         3 &        0.6 &      0.394 &       -0.5 &     -0.484 &        0.5 &      0.472 &        0.5 &      0.361 \\

         3 &        0.8 &      0.545 &       -0.5 &     -0.466 &        0.5 &      0.446 &        0.5 &      0.214 \\

\hline
\end{tabular}
\caption{\label{HR-nb-bias}Limiting HRMLE for  MVN copula-based models with marginal NB2 regression. The truncation point is 10.}
\end{center}
\end{small}
\end{table}

After  evaluating the adequacy of the MF and SL log-likelihood on finding  the peak (MLE), we evaluate if the curvature (Hessian) is also correct for the cases where the HRMLE and MSLE are correct. To check this, we also computed the negative inverse Hessian $H$ of the limit of the surrogate log-likelihood  in (\ref{limitlik}) and the  simulated log-likelihood in (\ref{limitslik}); because these are limits as $n\to\infty$ of $n^{-1}$ times the log-likelihood, $H$ is the inverse Fisher information, or equivalently, the
covariance matrix for sample size $n$ is approximately $n^{-1}H$.  For a comparison, we have also calculated the Hessian at the limit for the standard MLE. For simpler comparisons, we convert to standard errors (SE), say for a sample size of $n = 100$ (that is, square roots of the
diagonals of the above matrices divided by $n$). Some representative results are given in Table \ref{HR-logistic-se} and Table \ref{HR-nb-se} for an MVN copula-based model with marginal logistic and NB2 regression, respectively, with the MSLE results omitted because they were again identical with MLE up to three or four decimal places.

\begin{table}[!h]
\begin{small}
\begin{center}
\begin{tabular}{cccccccccc}
\hline
$d$&   $\rho$ &         $\beta_0$ &       $\beta_1$ & ML & HR & ML & HR & ML & HR \\
&&&&\multicolumn{2}{c}{SE($\hat\beta_0$)}&\multicolumn{2}{c}{SE($\hat\beta_1$)}&\multicolumn{2}{c}{SE($\hat\rho$)}\\\hline

         2 &        0.3 &       -0.5 &        0.5 &       0.16 &       0.15 &       0.22 &       0.21 &       0.11 &       0.07 \\

         2 &        0.6 &       -0.5 &        0.5 &       0.17 &       0.16 &       0.24 &       0.22 &       0.08 &       0.06 \\

         2 &        0.8 &       -0.5 &        0.5 &       0.18 &       0.16 &       0.26 &       0.22 &       0.05 &       0.06 \\
\hline
         5 &        0.3 &       -0.5 &        0.5 &       0.12 &       0.10 &       0.17 &       0.14 &       0.05 &       0.03 \\

         5 &        0.6 &       -0.5 &        0.5 &       0.15 &       0.11 &       0.21 &       0.16 &       0.05 &       0.03 \\

         5 &        0.8 &       -0.5 &        0.5 &       0.17 &       0.12 &       0.23 &       0.17 &       0.03 &       0.03 \\
\hline
        10 &        0.3 &       -0.5 &        0.5 &       0.11 &       0.08 &       0.15 &       0.11 &       0.03 &       0.02 \\

        10 &        0.6 &       -0.5 &        0.5 &       0.14 &       0.09 &       0.19 &       0.12 &       0.04 &       0.02 \\

        10 &        0.8 &       -0.5 &        0.5 &       0.16 &       0.09 &       0.22 &       0.13 &       0.03 &       0.02 \\

\hline
\end{tabular}

\caption{\label{HR-logistic-se}Standard errors (SE) of the limiting HRMLE and MLE for  MVN copula-based models with marginal logistic regression.}
\end{center}
\end{small}
\end{table}

\begin{table}[!h]
\begin{small}
\begin{center}
\begin{tabular}{ccccccccccccc}
\hline
$d$&   $\rho$ &         $\beta_0$ &       $\beta_1$ & $\gamma$& ML & HR & ML & HR & ML & HR & ML & HR\\
&&&&&\multicolumn{2}{c}{SE($\hat\beta_0$)}&\multicolumn{2}{c}{SE($\hat\beta_1$)}&\multicolumn{2}{c}{SE($\hat\gamma$)}&\multicolumn{2}{c}{SE($\hat\rho$)}\\\hline

         2 &        0.3 &       -0.5 &        0.5 &        0.5 &       0.11 &       0.11 &       0.15 &       0.14 &       0.15 &       0.14 &       0.08 &       0.07 \\

         2 &        0.6 &       -0.5 &        0.5 &        0.5 &       0.13 &       0.11 &       0.16 &       0.15 &       0.15 &       0.13 &       0.06 &       0.06 \\

         2 &        0.8 &       -0.5 &        0.5 &        0.5 &       0.13 &       0.12 &       0.17 &       0.15 &       0.17 &       0.12 &       0.04 &       0.04 \\
\hline
         3 &        0.3 &       -0.5 &        0.5 &        0.5 &       0.10 &       0.09 &       0.13 &       0.12 &       0.12 &       0.11 &       0.06 &       0.04 \\

         3 &        0.6 &       -0.5 &        0.5 &        0.5 &       0.12 &       0.10 &       0.15 &       0.13 &       0.13 &       0.10 &       0.05 &       0.04 \\

         3 &        0.8 &       -0.5 &        0.5 &        0.5 &       0.13 &       0.10 &       0.16 &       0.13 &       0.15 &       0.09 &       0.03 &       0.03 \\

\hline
\end{tabular}

\caption{\label{HR-nb-se}Standard errors (SE) of the limiting HRMLE and MLE for MVN copula-based models with marginal NB2 regression. The truncation point is 10.}
\end{center}
\end{small}
\end{table}

The results in Tables \ref{HR-logistic-se} and \ref{HR-nb-se} show that the HR method slightly underestimates the SE for the regression parameters. Underestimation of the curvature increases as  the dimension $d$ and/or the latent correlation $\rho$ increases. The HR method leads to underestimation of the SE, because it is using information on jitters that are not in the observed data.

To sum up, the asymptotics show that the maximum SL method is as good as maximum likelihood, while  the maximum surrogate log-likelihood using the CE leads to approximate asymptotic unbiasedness for (some) univariate marginal parameters and not for the latent correlation parameters, because the jittering is univariate and does not account for dependence.
Although we show the details only for exchangeable dependence, we expect the above results to hold in general, as well as to apply to different dependence structures. Section 5 contains small sample efficiency calculations using both exchangeable and AR(1) dependence.

Closing this section, we explain why the the surrogate log-likelihood using the CE fails. It is easiest notationally to indicate what is happening with bivariate discretized normal:
    $$Y_1=j_1,  Y_2=j_2 \quad \mbox{iff} \quad z_{1,j_1-1}<Z_1\leq z_{1,j_1},   \quad  z_{2,j_2-1}<Z_2\leq z_{2,j_2}$$
where $z_{1,j},z_{2,j}$ are cutpoints, and $(Z_1,Z_2)$ is bivariate standard normal with correlation $\rho.$
Then conditioned on $\{Y_1=j_1, Y_2=j_2\}$, $(Z_1,Z_2)$ is dependent with a truncated bivariate normal distribution.
This means the jittered variables $(V_1,V_2)$ have to be dependent to get the correct conditional distribution.
So, for jittering to be asymptotically unbiased, a sequential approach would be needed based on the previous estimates of the latent correlations.

\subsection{\label{MF-theoretical}Computation of MVN rectangle probabilities}
In this section, we describe how the high-dimensional  multivariate normal rectangle probability is computed by \cite{madsen09} and \cite{Madsen&Fang11} using importance sampling, and compare it with the naive simulation method, and the method in \cite{genz&bretz02}.

For an integral in the context of an
MVN rectangle probability in (\ref{rect-formula}),
naive simulation uses,
  $$m^{-1} \sum \mathbf 1(a_j<z_{j}< b_j, j=1,...,d) ,$$
where   $\mathbf 1(A)$ denotes the indicator function of the set $A$ and  $z_1,..., z_d$ are $m$ iid variates from $\phi_\R$.
Importance sampling gives,
   $$\int_{a_1}^{b_1}\cdots\int_{a_d}^{b_d} \bigl[\phi_\R(z_1,\ldots,z_d)/g(z_1,\ldots,z_d)\bigr] g(z_1,\ldots,z_d) dz_1\ldots dz_d,$$
where $g$ is a closed-form density from which it is easy to simulate.

Estimation is via,
 $$ m^{-1} \sum \phi_\R(z_1,\ldots,z_d)/g(z_1,\ldots,z_d),$$ where $z_1,..., z_d$ are iid simulated from $g$.
Better choices of $g$ have small variance for $w=\phi_\R/g$, such that $w$ is bounded (in which case one can bound the variance). The penalty for a bad $g$ can be longer run times than for a general Monte Carlo simulation without importance sampling.

\cite{madsen09} and \cite{Madsen&Fang11} implement  the  integration in (\ref{rect-formula}) by
transforming $z_1=\Phi^{-1}(\omega_1),\ldots, z_d=\Phi^{-1}(\omega_d)$ to get,
  $$\int_{\Phi(a_1)}^{\Phi(b_1)}\cdots \int_{\Phi(a_d)}^{\Phi(b_d)}
  \frac{\phi_\R\bigl(\Phi^{-1}(\omega_1),\ldots,\Phi^{-1}(\omega_d)\bigr)}
  {\phi\bigl(\Phi^{-1}(\omega_1)\bigr)\cdots \phi\bigr(\Phi^{-1}(\omega_d)\bigl) } \, d\omega_1 \ldots d\omega_d,$$
where $\phi$ is the standard normal density.
The \cite{denuit&lambert05} uniform extension in this case corresponds to evaluating the above integral based on a $d$-variate  uniform random sample  $\Obf_k=(\Omega_{1,k},\ldots,\Omega_{d,k})$, where $\Omega_{j,k},\,j=1,\ldots,d$ are uniform  in the interval $\Phi(a_j)$ to $\Phi(b_j)$ for $j=1,\ldots,d.$
Approximation is via,
$$
m^{-1}\sum_{k=1}^m \frac{\phi_\R\bigl(\Phi^{-1}(\omega_{1,k}),\ldots,\Phi^{-1}(\omega_{d,k})\bigr)\prod_{j=1}^d\bigl(\Phi(b_j)-\Phi(a_j)\bigr)}
  {\phi\bigl(\Phi^{-1}(\omega_{1,k})\bigr)\cdots \phi\bigr(\Phi^{-1}(\omega_{d,k})\bigl) },
$$
where $\obf_k=(\omega_{1,k},\ldots,\omega_{d,k})$ are realizations of the jitters $\Obf_k$.

\begin{table}[!h]
\begin{center}
\begin{tabular}{llllllllll}
\hline
         $d$ &          $a$ &        $\rho$ &      GB  & MF  &   SD & MF &   SD &      Naive &   SD \\
         &&&
         &\multicolumn{2}{c}{$m=10^3$}&$m=10^4$&&$m=10^4$\\\hline

         5 &          1 &        0.3 &      0.176 &      0.176 &      0.001 &      0.176 &      $<10^{-3}$ &      0.176 &      0.004 \\

             &                &        0.6 &      0.266 &      0.266 &      0.005 &      0.267 &      0.001 &      0.267 &      0.004 \\

             &                &        0.8 &      0.391 &      0.382 &      0.014 &      0.395 &      0.005 &      0.393 &      0.005 \\

             &          2 &        0.3 &      0.808 &      0.823 &      0.023 &      0.809 &      0.006 &      0.809 &      0.004 \\

             &                &        0.6 &      0.847 &      0.863 &      0.064 &      0.850 &      0.016 &      0.847 &      0.004 \\

             &                &        0.8 &      0.883 &      0.808 &      0.109 &      0.862 &      0.033 &      0.883 &      0.003 \\

             &          4 &        0.3 &      1.000 &      1.049 &      0.055 &      0.996 &      0.012 &      1.000 &      $<10^{-3}$ \\

            &                &        0.6 &      1.000 &      1.025 &      0.133 &      0.981 &      0.032 &      1.000 &      $<10^{-3}$ \\

             &                &        0.8 &      1.000 &      0.805 &      0.112 &      0.915 &      0.044 &      1.000 &      $<10^{-3}$ \\\hline

         10 &          1 &        0.3 &      0.038 &      0.037 &      $<10^{-3}$ &      0.038 &      $<10^{-3}$ &      0.039 &      0.002 \\

             &                &        0.6 &      0.110 &      0.107 &      0.003 &      0.111 &      0.001 &      0.113 &      0.003 \\

             &                &        0.8 &      0.267 &      0.244 &      0.020 &      0.270 &      0.007 &      0.270 &      0.004 \\

             &          2 &        0.3 &      0.674 &      0.641 &      0.040 &      0.677 &      0.010 &      0.675 &      0.005 \\

             &                &        0.6 &      0.768 &      0.741 &      0.192 &      0.753 &      0.033 &      0.763 &      0.004 \\

             &                &        0.8 &      0.840 &      0.647 &      0.293 &      0.659 &      0.063 &      0.837 &      0.004 \\

             &          4 &        0.3 &      0.999 &      0.912 &      0.111 &      0.972 &      0.024 &      1.000 &      $<10^{-3}$ \\

             &                &        0.6 &      0.999 &      1.001 &      0.420 &      0.890 &      0.057 &      1.000 &      $<10^{-3}$ \\

             &                &        0.8 &      1.000 &      0.621 &      0.289 &      0.624 &      0.069 &      1.000 &      $<10^{-3}$ \\
    \hline
        20 &          1 &        0.3 &      0.002 &      0.002 &      $<10^{-3}$ &      0.002 &      $<10^{-3}$ &      0.002 &      $<10^{-3}$ \\

            &                &        0.6 &      0.024 &      0.023 &      0.001 &      0.025 &      $<10^{-3}$ &      0.025 &      0.002 \\

             &                &        0.8 &      0.156 &      0.121 &      0.024 &      0.168 &      0.010 &      0.159 &      0.004 \\

             &          2 &        0.3 &      0.493 &      0.422 &      0.016 &      0.498 &      0.019 &      0.496 &      0.005 \\

             &                &        0.6 &      0.670 &      0.354 &      0.064 &      0.633 &      0.047 &      0.671 &      0.005 \\

             &                &        0.8 &      0.792 &      0.478 &      0.451 &      0.673 &      0.283 &      0.792 &      0.004 \\

             &          4 &        0.3 &      0.999 &      0.672 &      0.036 &      0.959 &      0.091 &      0.998 &      $<10^{-3}$ \\

             &                &        0.6 &      0.999 &      0.347 &      0.107 &      0.748 &      0.073 &      0.999 &      $<10^{-3}$ \\

             &                &        0.8 &      0.999 &      0.514 &      0.503 &      0.699 &      0.341 &      0.999 &      $<10^{-3}$ \\

\hline
\end{tabular}
\caption{\label{MF-table}Comparisons of the accuracy of the MF, the naive method, and the method in Genz and Bretz (GB, 2002) for the equicorrelated rectangle probability of the form $\Pr(-a\leq Z_j\leq a, j=1,\ldots,d)$    for dimensions $d=5,10,20$ and correlations $\rho=0.3,0.6,0.8$. The computed probabilities with the GB method are identical up to three or four decimal places with the ones found by the numerically accurate method in \cite{Johnson&Kotz72}.}
\end{center}
\end{table}

In Table \ref{MF-table}, comparisons of the accuracy of the MF ($m=10^3$ and $m=10^4$), the naive ($m=10^4$) method, and the method in \cite{genz&bretz02} are presented. The accuracy comparisons are for the computation of the equicorrelated rectangle probability of the form $\Pr(-a\leq Z_j\leq a, j=1,\ldots,d)$ for dimensions $d=5,10,20$ and correlations $\rho=0.3,0.6,0.8$. These probabilities  are also computed with the numerically accurate method in \cite{Johnson&Kotz72},
and  the results are identical up to three or four decimal places with the ones found by the method in \cite{genz&bretz02}.
For these MVN rectangle probabilities,  we simulate based on the
MF  method, keeping  track of the values of $w(\cdot)$ when simulating from
the $d$-variate  uniform   random sample. Based on the sample variance of $w(\cdot)$, we estimate the
achieved  accuracy  at the number of replications $m$,
that is, SD$=\sqrt{Var[w(\cdot)]/m} <$ accuracy. This calculation is simple for the naive method.

In Table \ref{MF-table}, it is clear that the MF method (even with $m=10^4$) gets worse as

\begin{enumerate}
\itemsep=0pt
\item the dimension $d$ increases;
\item  the latent correlation $\rho$ increases;
\item the limits (integrated region) increase;
\end{enumerate}
and even the naive method is much better; it has 2 decimal place accuracy for $m=10^4$ replications. The use of jitters with  $m=10^3$,  as in \cite{Madsen&Fang11}, is highly inefficient even in a low dimension.
Therefore, the MF method is a very inefficient way to compute a multivariate normal rectangle probability, which means that even  $m=10^4$  is far from sufficient to achieve a desired accuracy. It is also shown to be difficult to even estimate the SD, because the integrand has no bound.

Regarding equation (\ref{MF-equation}),  this is much worse than approximating many $d$-dimensional MVN rectangle probabilities separately, that is, separate jitters of each $d$-dimensional probability.
This is actually a worse way to do simulated likelihood compared with what we mention above.

\hyphenation{par-am-etr-iz-at-ion}

\section{Simulations}

In this section we performed several simulation studies to assess  the performance of the HR, MF, and SL methods.
We used structured latent correlation matrices for the MVN copula. For exchangeable dependence, we took
$\R$ as $(1-\rho)\I_d+ \rho \J_d$,
where $\I_d$ is the identity matrix of order $d$ and $\J_d$ is the
$d\times d$ matrix of 1s. For  AR(1) dependence, $\R$ was taken
as $(\rho^{|j-k|})_{1\le j,k\le d}$.

For marginal models we used Bernoulli$(\mu)$, Poisson$(\mu)$,  NB$1(\mu,\,\ga)$,  and NB$2(\mu,\,\ga)$ parametrization  of the negative binomial distribution. For the covariates and regression parameters,
we chose $p=2, \x_j=(\mathbf 1,x_j)^\top,$ where $x_j$ drawn from a ${U}[-1, 1]$, and let $\mu$ depend on the covariates, that is $\eta(\mu_j)=(\beta_0 + \beta_1 x_j), \,j=1,\ldots,d,$ where $\beta_1=-\beta_0=0.5.$ For the link function $\eta$, we took the log link function for Poisson and NB regression, and the logit link function or the probit link function for binary regression. Note also that binary and Poisson regression $\gbf$ is null, while for NB1 and NB2 regression $\gamma$ is scalar ($r=1$).

\subsection{Assessing the variability due only to jittering}
As a first step, we assess the variability of the HR and MF estimates over different sets of jitters $\mathbf V_k,\,k=1,\ldots,m$.
Our goal is to define the value of $m$ for which different sets of jitters for the same data set reproduce the same results.
To this end, we fixed one simulated dataset and used 5 sets of random jitters $\mathbf V_k,\,k=1,\ldots,m$; the number  of jitters $m$  was set at
$m = 10, 10^2, 10^3, 10^4.$
The following experiments are typical of the results that were obtained.

\begin{table}[!h]
\begin{center}
\begin{tabular}{ccccccc}
\hline
       &  \multicolumn{3}{c}{MF $(m=10)$}     &   \multicolumn{3}{c}{HR $(m=10)$}  \\
    set   & $\hat\beta_0$ &  $\hat\beta_1$ &  $\hat\rho$ & $\hat\beta_0$ &  $\hat\beta_1$ &  $\hat\rho$\\
\hline
         1 &      -0.51 &       0.40 &       0.20 &      -0.51 &       0.40 &       0.15 \\

         2 &      -0.54 &       0.44 &       0.33 &      -0.51 &       0.42 &       0.18 \\

         3 &      -0.52 &       0.41 &       0.27 &      -0.51 &       0.41 &       0.16 \\

         4 &      -0.50 &       0.45 &       0.27 &      -0.51 &       0.42 &       0.16 \\

         5 &      -0.51 &       0.41 &       0.25 &      -0.51 &       0.41 &       0.21 \\
\hline
       &  \multicolumn{3}{c}{MF $(m=10^2)$}     &   \multicolumn{3}{c}{HR $(m=10^2)$}  \\
    set   & $\hat\beta_0$ &  $\hat\beta_1$ &  $\hat\rho$ & $\hat\beta_0$ &  $\hat\beta_1$ &  $\hat\rho$\\
\hline

         1 &      -0.52 &       0.33 &       0.39 &      -0.51 &       0.41 &       0.16 \\

         2 &      -0.52 &       0.45 &       0.32 &      -0.51 &       0.42 &       0.18 \\

         3 &      -0.52 &       0.47 &       0.39 &      -0.51 &       0.42 &       0.18 \\

         4 &      -0.52 &       0.41 &       0.33 &      -0.51 &       0.42 &       0.17 \\

         5 &      -0.49 &       0.46 &       0.30 &      -0.51 &       0.42 &       0.17 \\
\hline
       &  \multicolumn{3}{c}{MF $(m=10^3)$}     &   \multicolumn{3}{c}{HR $(m=10^3)$}  \\
    set   & $\hat\beta_0$ &  $\hat\beta_1$ &  $\hat\rho$ & $\hat\beta_0$ &  $\hat\beta_1$ &  $\hat\rho$\\
\hline

         1 &      -0.52 &       0.44 &       0.35 &      -0.51 &       0.42 &       0.17 \\

         2 &      -0.51 &       0.45 &       0.38 &      -0.51 &       0.42 &       0.17 \\

         3 &      -0.51 &       0.45 &       0.36 &      -0.51 &       0.42 &       0.17 \\

         4 &      -0.49 &       0.49 &       0.42 &      -0.51 &       0.42 &       0.17 \\

         5 &      -0.50 &       0.47 &       0.32 &      -0.51 &       0.42 &       0.17 \\
\hline
       &  \multicolumn{3}{c}{MF $(m=10^4)$}     &   \multicolumn{3}{c}{HR $(m=10^4)$}  \\
    set   & $\hat\beta_0$ &  $\hat\beta_1$ &  $\hat\rho$ & $\hat\beta_0$ &  $\hat\beta_1$ &  $\hat\rho$\\
\hline
         1 &      -0.51 &       0.45 &       0.36 &      -0.51 &       0.42 &       0.17 \\

         2 &      -0.52 &       0.46 &       0.38 &      -0.51 &       0.42 &       0.17 \\

         3 &      -0.51 &       0.45 &       0.38 &      -0.51 &       0.42 &       0.17 \\

         4 &      -0.50 &       0.46 &       0.37 &      -0.51 &       0.42 &       0.17 \\

         5 &      -0.51 &       0.45 &       0.34 &      -0.51 &       0.42 &       0.17 \\
\hline
ML & $\hat\beta_0$ & SE& $\hat\beta_1$ &  SE& $\hat\rho$ &SE\\
           &      -0.51 &       0.03 &       0.47 &       0.07 &       0.43 &       0.02 \\
\hline
\end{tabular}
\caption{\label{select1}Variability of the MF and HR estimates with different sets of jitters $\mathbf V_k,\,k=1,\ldots,m$, along  with ML estimates and standard errors (SE) for
a fixed simulated data set with $n=100$ observations from
the bivariate normal copula
model with moderate dependence $(\rho=0.5)$ and logistic regression. The
number  of jitters  was set at
$m = 10, 10^2, 10^3, 10^4.$}
\end{center}
\end{table}

We fixed one simulated data set with $n=100$ observations from the bivariate normal copula with exchangeable moderate dependence ($\rho=0.5$) and marginal logistic regression. Table \ref{select1} shows the variability of the HR and MF estimators over 5 different sets of jitters as compared with ML estimates and standard errors; for $d=2$ there is a numerical accurate calculation of the bivariate normal rectangle probabilities in (\ref{MLlik}). To investigate if these results hold for higher dimensions,
we also fixed one simulated data set with $n=100$ observations from the MVN $(d=5)$ copula  with strong AR(1) dependence ($\rho=0.8$), and NB2 regression with large overdispersion $\ga=2$.
Table \ref{select2} shows the variability of the HR and MF estimators over 5 different sets of jitters as compared with maximum SL estimates and standard errors. These results and other computations that we have done  with latent correlation $\rho$ varying from 0 to 0.9 in 0.1 increments confirm that  for the HR (MF) method, $m=100$ ($m=1000$) is sufficient  to produce  the same estimates over different sets of jitters.
Note that the SEs of the ML and maximum SL estimates  are obtained via the square roots of the diagonals of the inverse Hessian computed
numerically during the maximization process.

\begin{table}[!h]
\begin{center}
\begin{tabular}{ccccccccc}
\hline
       &  \multicolumn{4}{c}{MF $(m=10)$}     &   \multicolumn{4}{c}{HR $(m=10)$}  \\
    set   & $\hat\beta_0$ &  $\hat\beta_1$ & $\hat\gamma$& $\hat\rho$ & $\hat\beta_0$ &  $\hat\beta_1$ & $\hat\gamma$&  $\hat\rho$\\
\hline
         1 &      -0.44 &       0.49 &       1.14 &       0.50 &      -0.42 &       0.48 &       1.14 &       0.50 \\

         2 &      -0.39 &       0.49 &       1.13 &       0.53 &      -0.42 &       0.46 &       1.14 &       0.50 \\

         3 &      -0.45 &       0.53 &       1.09 &       0.50 &      -0.43 &       0.50 &       1.11 &       0.48 \\

         4 &      -0.43 &       0.42 &       1.16 &       0.53 &      -0.42 &       0.48 &       1.15 &       0.50 \\

         5 &      -0.39 &       0.56 &       1.08 &       0.52 &      -0.42 &       0.50 &       1.14 &       0.49 \\
\hline
       &  \multicolumn{4}{c}{MF $(m=10^2)$}     &   \multicolumn{4}{c}{HR $(m=10^2)$}  \\
    set   & $\hat\beta_0$ &  $\hat\beta_1$ & $\hat\gamma$& $\hat\rho$ & $\hat\beta_0$ &  $\hat\beta_1$ & $\hat\gamma$&  $\hat\rho$\\
\hline
         1 &      -0.37 &       0.54 &       1.13 &       0.58 &      -0.42 &       0.49 &       1.13 &       0.49 \\

         2 &      -0.40 &       0.47 &       1.20 &       0.57 &      -0.42 &       0.49 &       1.12 &       0.50 \\

         3 &      -0.41 &       0.47 &       1.17 &       0.56 &      -0.42 &       0.48 &       1.12 &       0.49 \\

         4 &      -0.42 &       0.50 &       1.11 &       0.53 &      -0.42 &       0.49 &       1.13 &       0.49 \\

         5 &      -0.43 &       0.51 &       1.17 &       0.55 &      -0.42 &       0.48 &       1.13 &       0.49 \\
\hline
       &  \multicolumn{4}{c}{MF $(m=10^3)$}     &   \multicolumn{4}{c}{HR $(m=10^3)$}  \\
    set   & $\hat\beta_0$ &  $\hat\beta_1$ & $\hat\gamma$& $\hat\rho$ & $\hat\beta_0$ &  $\hat\beta_1$ & $\hat\gamma$&  $\hat\rho$\\
\hline
         1 &      -0.38 &       0.48 &       1.13 &       0.58 &      -0.42 &       0.48 &       1.13 &       0.49 \\

         2 &      -0.38 &       0.46 &       1.24 &       0.61 &      -0.42 &       0.48 &       1.13 &       0.49 \\

         3 &      -0.37 &       0.48 &       1.16 &       0.58 &      -0.42 &       0.48 &       1.13 &       0.49 \\

         4 &      -0.39 &       0.48 &       1.20 &       0.57 &      -0.42 &       0.49 &       1.12 &       0.49 \\

         5 &      -0.41 &       0.49 &       1.16 &       0.56 &      -0.42 &       0.48 &       1.13 &       0.49 \\
\hline
SL & $\hat\beta_0$ & SE& $\hat\beta_1$ &  SE& $\hat\gamma$& SE& $\hat\rho$ &SE\\
&     -0.37 &       0.15 &       0.49 &       0.09 &       1.96 &       0.40 &       0.80 &       0.03 \\
\hline
\end{tabular}
\caption{\label{select2}Variability of the MF and HR estimates with different sets of jitters $\mathbf V_k,\,k=1,\ldots,m$, along  with maximum SL estimates and standard errors (SE)
for a fixed simulated data set with $n=100$ observations from  the MVN $(d=5)$ copula
model with strong dependence $(\rho=0.8)$ and NB2 regression. The
number  of jitters  was set at
$m = 10, 10^2, 10^3.$}
\end{center}
\end{table}

\subsection{Small-sample efficiency}
In this subsection, we gauge the small-sample efficiency  of the HR, MF, and SL methods.  The number of jitters  was set at
$m = 10, 10^2, 10^3, 10^4,$ and the sample size at $n=100, 300, 500$. The following experiments are typical of the results that were obtained. We report simulations from small sample sizes ($n=100$), since the additional simulation results do not show sensitivity to $n$. Note that the theoretical variances of the maximum SL and MF estimates are obtained via the gradients and the Hessian computed
numerically during the maximization process, while the HR variances are obtained by averaging the latter  over $m$  runs.

For $n=100$, $10^4$ random samples of size $n$ were generated from the bivariate normal copula with exchangeable moderate dependence ($\rho=0.5$), and marginal logistic regression.
Table~\ref{biasmse} contains the parameter values, the bias,
variance (Var), and mean square errors (MSE) of the
MF, HR and ML (for $d=2$ there is a numerical accurate calculation of the bivariate probabilities) estimates,  along
with the average of their theoretical variances ($\bar V$).

\begin{table}[!h]
\begin{center}
 \begin{tabular}{ccccccc}
\hline    Method &            &          $m$ &      $n$Bias &       $n$Var &       $n$MSE &         $n\bar V$ \\
\hline
        MF &     &         $10$ &      -0.56 &       3.02 &       3.02 &       2.61 \\

           &            &        $10^2$ &      -0.54 &       3.02 &       3.02 &       2.69 \\

           &            &       $10^3$ &      -0.58 &       3.03 &       3.03 &       2.75 \\
           &&$10^4$& -0.63	&3.02	&3.03	&2.80\\

        HR &    $\beta_0=-0.5$        &         $10$ &      -0.60 &       3.00 &       3.00 &        2.48
    \\

           &            &        $10^2$ &      -0.60 &       2.99 &       3.00 &         2.48
   \\

           &            &       $10^3$ &      -0.60 &       2.99 &       2.99 &        2.48
    \\
           && $10^4$	&-0.60	&2.99	&2.99&2.47
\\

        ML &            &           &      -0.73 &       3.00 &       3.00 &       2.92 \\
\hline
        MF &     &         $10$ &       0.70 &       6.54 &       6.54 &       6.56 \\

           &            &        $10^2$ &       0.76 &       6.49 &       6.50 &       6.48 \\

           &            &       $10^3$ &       0.83 &       6.45 &       6.45 &       6.41 \\
           &&
           $10^4$	&0.82	&6.41	&6.42	&6.35\\

        HR &     $\beta_1=0.5$        &         $10$ &       0.70 &       6.57 &       6.58 &          6.67   \\

           &            &        $10^2$ &       0.70 &       6.57 &       6.57 &        6.67    \\

           &            &       $10^3$ &       0.70 &       6.56 &       6.57 &         6.67    \\
           &&$10^4$	&0.70	&6.56	&6.57	& 6.67\\

        ML &            &            &       0.85 &       6.31 &       6.31 &       6.15 \\
\hline
        MF &     &         $10$ &     -20.93 &       0.77 &       5.15 &       0.96 \\

           &            &        $10^2$ &     -15.17 &       0.74 &       3.05 &       1.00 \\

           &            &       $10^3$ &     -11.19 &       0.78 &       2.03 &       1.06 \\
           && $10^4$	&-8.30	&0.87	&1.56	&1.12\\

        HR &    $\rho=0.5$        &         $10$ &     -30.42 &       0.42 &       9.67 &         0.90   \\

           &            &        $10^2$ &     -30.40 &       0.37 &       9.61 &         0.90   \\

           &            &       $10^3$ &     -30.39 &       0.36 &       9.59 &         0.90   \\
           && $10^4$	&-30.39&	0.36&	9.59&	0.90\\

        ML &            &           &      -0.73 &       1.84 &       1.84 &       1.82 \\
\hline

\end{tabular}
\caption{\label{biasmse} Small sample of sizes $n=100$ simulations ($10^4$ replications) and resultant biases and mean square errors (MSE) and variances (Var), along with average theoretical variances scaled by $n$, for the
MF, HR, and ML of the regression and copula parameters for the bivariate normal copula
model with moderate dependence and logistic regression. The
number  of jitters  was set at
$m = 10, 10^2, 10^3, 10^4.$}
\end{center}
\end{table}

\begin{table}[!h]
\begin{center}
 \begin{tabular}{cclcccc}
\hline    Method &            &          $m$ &      $n$Bias &       $n$Var &       $n$MSE &         $n\bar V$ \\
\hline

        MF &            &         $10$ &      -5.94 &       2.48 &       2.84 &       1.10 \\

           &            &        $10^2$ &      -5.89 &       2.49 &       2.83 &       1.15 \\

           &            &       $10^3$ &      -5.82 &       2.50 &       2.84 &       1.19 \\

        HR &    $\beta_0=-0.5$ &         $10$ &      -5.88 &       2.43 &       2.78 &       1.03 \\

           &            &        $10^2$ &      -5.89 &       2.43 &       2.77 &       1.04 \\

           &            &       $10^3$ &      -5.88 &       2.43 &       2.77 &       1.04 \\

               SL &     &            &      -1.36 &       2.33 &       2.35 &       2.28 \\
\hline
        MF &            &         $10$ &      -0.72 &       1.09 &       1.10 &       1.32 \\

           &            &        $10^2$ &      -0.55 &       1.07 &       1.08 &       1.28 \\

           &            &       $10^3$ &      -0.35 &       1.05 &       1.05 &       1.26 \\

        HR &     $\beta_1=0.5$ &         $10$ &      -1.06 &       1.03 &       1.04 &       1.39 \\

           &            &        $10^2$ &      -1.07 &       1.02 &       1.03 &       1.39 \\

           &            &       $10^3$ &      -1.07 &       1.02 &       1.03 &       1.39 \\

     SL &     &            &       0.12 &       0.89 &       0.89 &       0.83 \\
\hline
        MF &            &         $10$ &     -78.13 &       8.36 &      69.41 &       6.51 \\

           &            &        $10^2$ &     -77.77 &       8.13 &      68.62 &       6.51 \\

           &            &       $10^3$ &     -77.18 &       8.01 &      67.57 &       6.54 \\

        HR &      $\gamma=2$ &         $10$ &     -77.27 &       8.71 &      68.42 &       6.62 \\

           &            &        $10^2$ &     -77.24 &       8.68 &      68.34 &       6.62 \\

           &            &       $10^3$ &     -77.24 &       8.68 &      68.33 &       6.62 \\

       SL &      &            &      -1.32 &      19.03 &      19.05 &      18.59 \\
\hline
        MF &            &         $10$ &     -28.70 &       0.20 &       8.44 &       0.12 \\

           &            &        $10^2$ &     -25.96 &       0.18 &       6.92 &       0.11 \\

           &            &       $10^3$ &     -23.95 &       0.16 &       5.90 &       0.11 \\

        HR &    $\rho=0.8$ &         $10$ &     -33.23 &       0.19 &      11.23 &       0.13 \\

           &            &        $10^2$ &     -33.22 &       0.18 &      11.21 &       0.13 \\

           &            &       $10^3$ &     -33.21 &       0.18 &      11.21 &       0.13 \\

       SL &    &            &       0.88 &       0.11 &       0.12 &       0.11 \\
\hline

\end{tabular}
\caption{\label{biasmsecl}Small sample of sizes $n=100$ simulations ($10^4$ replications) and resultant biases and mean square errors (MSE) and variances (Var),
along with average theoretical variances scaled by $n$, for the
HR, MF, and SL of the regression and copula parameters for the MVN $(d=5)$ copula
model with strong dependence and NB2 regression. The
number of jitters was set at
$m = 10, 10^2, 10^3.$}
\end{center}
\end{table}

For $n=100$, $10^4$ random samples of size $n$ were generated from the MVN $(d=5)$ copula with strong AR(1) dependence ($\rho=0.8$)
and NB2 regression with large overdispersion $\gamma=2.$ Table~\ref{biasmsecl} contains the parameter values, the bias,
variance (Var) and mean square errors (MSE) of the
MF, HR and maximum SL estimates,  along
with the average of their theoretical variances ($\bar V$).
Because HR and MF likelihood takes much longer with larger $m$ and $d$, we
ran fewer replications on a subset for the 5-dimensional case. With an Intel Core Duo 2$\times$ 2.53Ghz
processor, the computing times
in an R program for SL, MF and HR with $m=10^3$ jitters, averaged about 1, 6, and 6 minutes respectively; the time is about 1 hour for HR and MF with $m=10^4$ jitters. To this end, simulations have been restricted to  $m=10^3$.

Conclusions from the values in the  tables
and other computations that we have done are the following:

\begin{enumerate}
\itemsep=-1pt
\item The SL method is highly efficient according to the simulated biases and variances.
\item The HR and MF methods yield estimates that are almost
as good as the ML and maximum SL estimates for the regression parameters.

\item The HR and MF methods  underestimate  the univariate marginal parameters that  are not regression coefficients as the latent correlation increases.

\item The efficiency of HR and MF methods  is low
for the latent correlation, for a wide range
of $\rho$ values. The dependence parameters (latent correlations) are underestimated.

\item For the MF method, it appears that there are improvements with $m=10^4$, so one might wonder if $m=10^6$ would be better. This number of jitters is prohibitive with respect to the computational time as the dimension increases. Further, theoretically, there are still problems; see subsection \ref{MF-theoretical}.

\item
Overall, efficiencies of HR do not change considerably, when $m$ changes.

\item The variances for the MF and HR estimates are underestimated for the non-regression parameters,  the intercept, and the regression parameters of discrete covariates.

\item The variances for the MF and HR estimates are overestimated for  the regression parameters of continuous covariates that do not vary a lot.

\end{enumerate}
These conclusions justify why \cite{madsen09} and \cite{Madsen&Fang11} do not report latent correlations
to their applications and studies of efficiency.

\section{The toenail infection data}
In this section we re-analyze the toenail infection data in \cite{Madsen&Fang11}.
The data were obtained from a randomized,
double-blind, parallel group, multicenter study for the comparison of two
oral treatments for toenail dermatophyte
onychomycosis \citep{DeBacker-etal-1996,Molenberghs&Verbeke2005}.  Subjects were followed during 12 weeks (3 months) of treatment and followed
further, up to a total of 48 weeks (12 months). Measurements were
taken at baseline, every month during treatment, and every 3 months afterwards,
resulting in a maximum of 7 measurements per subject.
The observations are coded as 1 if
the subject's infection was severe and 0 otherwise. The question of interest was whether the percentage of severe infections
decreased over time, and whether that evolution was different for the two
treatment groups.
In accordance with \cite{Madsen&Fang11},  we use the 224 subjects observed at all
seven time points, though the SL method does not depend on a constant ``cluster" size $d$.
In this data example,  \cite{Madsen&Fang11} assumed an exchangeable structure for the MVN copula with logistic regressions, and hence this can be easily analyzed with the standard ML method. ML estimation  is possible for an exchangeable structure, since the $7$-dimensional rectangle reduces to 1-dimensional integral.

\begin{table}[!h]
\begin{center}
\begin{tabular}{ccccccccc}
\hline
Dependence           &             \multicolumn{ 4}{c}{AR(1) } &            \multicolumn{ 4}{c}{Markov } \\

Link           & \multicolumn{ 2}{c}{probit} & \multicolumn{ 2}{c}{logit} & \multicolumn{ 2}{c}{probit} & \multicolumn{ 2}{c}{logit} \\
\hline
           &       Est. &         SE &       Est. &         SE &       Est. &         SE &       Est. &         SE \\
\hline
 Intercept &     -0.418 &      0.116 &     -0.642 &      0.189 &     -0.385 &      0.119 &     -0.587 &      0.194 \\

 Treatment &     -0.006 &      0.160 &      0.009 &      0.262 &     -0.011 &      0.165 &     -0.006 &      0.268 \\

      Time &     -0.099 &      0.016 &     -0.188 &      0.032 &     -0.111 &      0.016 &     -0.208 &      0.033 \\

  Trt.$\times$time &     -0.022 &      0.023 &     -0.054 &      0.048 &     -0.021 &      0.024 &     -0.048 &      0.048 \\

       $\rho$ &      0.923 &      0.015 &      0.924 &      0.015 &      0.952 &      0.010 &      0.952 &      0.010 \\
\hline
       $\ell$ & \multicolumn{ 2}{c}{-412.58} & \multicolumn{ 2}{c}{-410.06} & \multicolumn{ 2}{c}{-405.71} & \multicolumn{ 2}{c}{-403.49} \\
\hline
\end{tabular}

\caption{\label{exar}Maximum
SL log-likelihoods ($\ell$'s), estimates and standard errors (SE) for the the toenail infection data.}
\end{center}

\end{table}

However, for longitudinal data, it is  common practice to use various  parametric correlation
matrices, so we have also calculated the estimates under an AR(1), and Markov dependence  with the SL method. Using a Markov model, the AR(1) model is generalized to times that are unequally-spaced, and this is the case for toenail infection data. For  Markov dependence, $\R$ is taken
as $(\rho^{|t_j-t_k|})_{1\le j,k\le d}$.
Further, both  logit and probit links are used for the marginal binary regressions.
In Table \ref{exar}, we report  the resulting maximum SL log-likelihoods ($\ell$'s), estimates and standard errors (SE)  of the MVN copula-based models with binary regression. The SL log-likelihoods show that  Markov dependence is marginally better than  AR(1) dependence, and both are far better than exchangeable dependence \citep[Table 1]{sabo&chaganty11}. Further, logistic regression is slightly better than probit regression.

Based on our analysis and the exchangeable analysis, the
standard errors show the time effect to be highly significant, and the treatment by time interaction
insignificant; hence there is no (significant) difference in evolution between both treatment
groups. However, for an AR(1) or Markov structure, the $p$-value for the treatment by time interaction is smaller than its counterpart for the exchangeable analysis. Generally speaking, this implies that ignoring the actual correlation structure in the data could lead to invalid conclusions, although this was not crucial in this example.

\section{Discussion}
In this paper we have studied two ``approximate" likelihood estimation methods based on the CE of a discrete random variable.
For the binary, Poisson, and negative binomial regression models with the MVN copula, we have shown that these methods lead to substantial downward bias for the estimates  of the latent correlation and the univariate marginal parameters that they are not regression coefficients; for the latter parameters only for strong dependence.

We have shown that the simulated likelihood in \cite{madsen09} and \cite{Madsen&Fang11}  is a very inefficient way to compute a multivariate normal rectangle probability, since the importance weight is not bounded, and even the naive method is much better. The inefficiency of their method also yields to the claim that the GEE \citep{liang&zeger86} is more efficient than the ML. A moment based estimate cannot be better than a maximum (simulated) likelihood estimate, and this has to do with the inefficiency of the MF method. See \cite{sabo&chaganty11} and \cite{Song11} for further criticism.

We have implemented  a simulated likelihood method, where the rectangles are converted to
bounded integrands via the method in \cite{genz&bretz02}, and hence  the computational and statistical efficiency of simulated likelihood is substantially improved, and actually is as good as maximum likelihood as shown for  dimension 10 or lower. We expect our findings to hold in higher dimensions. Although there is an issue of computational burden as the dimension and the sample size increase,  this will become marginal, as computing technology is advancing rapidly.

\section*{Appendix. Derivation of $\xi_j^{(t)}$ and $\zeta_j^{(t)}$}
Split the $n$ observations into the distinct sets $n^{(t)}=\{i:\y_i=\y^{(t)},\x_i=\x^{(t)}\}.$
As $n\to\infty,$ then  with convergence in probability,

\begin{eqnarray*}n^{-1}\sum_{i\in n^{(t)}}q_{ij}^2&=&n^{-1}\sum_{i\in n^{(t)}}[\Phi^{-1}\{F_{Y_j}(y_{ij}-1;\x_{ij}) + v_{ij} f_{Y_j}(y_{ij};\x_{ij})\}]^2\\
&\to& p^{(t)}\int_0^1[\Phi^{-1}\{F_{Y_j}(y_{j}^{(t)}-1;\x_j^{(t)}) + v f_{Y_j}(y_{j}^{(t)};\x_j^{(t)})\}]^2 dv\\
&:=&p^{(t)}\xi_{j}^{(t)}
\end{eqnarray*}

\begin{center} and \end{center}

\begin{eqnarray*}n^{-1}\sum_{i\in n^{(t)}}q_{ij}q_{ik}&=&n^{-1}\sum_{i\in n^{(t)}}\Phi^{-1}\{F_{Y_j}(y_{ij}-1;\x_{ij}) + v_{ij} f_{Y_j}(y_{ij};\x_{ij})\}\times\\
&&\qquad \qquad \Phi^{-1}\{F_{Y_j}(y_{ik}-1;\x_{ik}) + v_{ik} f_{Y_j}(y_{ik};\x_{ik})\}\\
&\to& p^{(t)}\int_0^1\Phi^{-1}\{F_{Y_j}(y_{j}^{(t)}-1;\x_j^{(t)}) + v_j f_{Y_j}(y_{j}^{(t)};\x_j^{(t)})\} dv_j\times\\
&&\quad  \int_0^1\Phi^{-1}\{F_{Y_k}(y_{k}^{(t)}-1;\x_k^{(t)}) + v_k f_{Y_k}(y_{k}^{(t)};\x_k^{(t)})\} dv_k\\
&:=&p^{(t)}\zeta_{j}^{(t)}\zeta_{k}^{(t)}.
\end{eqnarray*}
Note that $\xi_{j}^{(t)},\zeta_{j}^{(t)},\,j=1,\ldots,d,t=1,\ldots,T$ are conditional expectations that have closed forms. For example with $Z\sim N(0,1)$,

\begin{eqnarray*}
\zeta_{j}^{(t)}&=&\int_0^1\Phi^{-1}\{F_{Y_j}(y_{j}^{(t)}-1;\x_j^{(t)}) + v_j f_{Y_j}(y_{j}^{(t)};\x_j^{(t)})\} dv_j\\
&=&\frac{1}{f_{Y_j}(y_{j}^{(t)};\x_j^{(t)})}\int_{\Phi^{-1}\{F_{Y_j}(y_{j}^{(t)}-1;\x_j^{(t)})\}}^{\Phi^{-1}
\{F_{Y_j}(y_{j}^{(t)};\x_j^{(t)})\}}z\phi(z)dz\\
&=&E\bigl[Z|\Phi^{-1}\{F_{Y_j}(y_{j}^{(t)}-1;\x_j^{(t)})\} \leq Z \leq \Phi^{-1}\{F_{Y_j}(y_{j}^{(t)};\x_j^{(t)})\}\bigr]
\end{eqnarray*}
\vspace{-5ex}
\begin{center} and \end{center}
\vspace{-1ex}
$$\xi_{j}^{(t)}=E\bigl[Z^2|\Phi^{-1}\{F_{Y_j}(y_{j}^{(t)}-1;\x_j^{(t)})\} \leq Z \leq \Phi^{-1}\{F_{Y_j}(y_{j}^{(t)};\x_j^{(t)})\}\bigr].$$

\section*{Acknowledgements}
Special thanks to Professor Harry Joe, University of British Columbia, for helpful  comments and suggestions.

\end{document}